\documentclass[twoside]{book}
\usepackage{graphicx}       
\usepackage{amsmath}
\usepackage{url}
\usepackage{epsfig}
\usepackage{setspace} 
\usepackage{array} 
\usepackage{multicol,threeparttable}

 \makeatletter
\ifx\UNDEF\mail\def\mail{ }\else\fi
\ifx\UNDEF\prange\def\prange{0 0}\else\fi

\gdef\@empty{}
\def\Mail#1 #2 {\gdef\thecontact{#1}\gdef\theaddr{#2}}
\def\Range#1 #2 {\gdef\thefirstpage{#1}\gdef\thelastpage{#2}}
{\let\'\mail \expandafter\Mail\' }  
{\let\'\prange \expandafter\Range\' }   
 \gdef\@shtitle{\relax}
 \long\def\shtitle#1{\gdef\@shtitle{#1}}
 \long\def\author#1{\gdef\@author{#1}}
 \def\affil#1{\par\noindent{\rm#1\par}}
 \gdef\@abstract{}
 \long\def\abstract#1{\gdef\@abstract{#1}}

 \renewcommand{\@evenhead}{\thepage\qquad\qquad\@shtitle\hfil}
 \renewcommand{\@oddhead}{\hfil\@shtitle\qquad\qquad\thepage}

 \def\maketitle{\thispagestyle{empty}\chapter{\@title}}
 \renewcommand\chapter{\if@openright\cleardoublepage\else\clearpage\fi
                    \thispagestyle{empty}%
                    \global\@topnum\z@
                    \@afterindentfalse
                    \secdef\@chapter\@schapter}
 \def\@makechapterhead#1{%
  {\parindent \z@ \raggedleft \normalfont
    \ifnum \c@secnumdepth >\m@ne
      \if@mainmatter
        \par\nobreak
        \vskip 20\p@
      \fi
    \fi
    \interlinepenalty\@M
    \huge \bfseries #1\par\nobreak
    \vskip.25in
    \large\bfseries\@author\par\nobreak
    \vskip 40\p@}
    \ifx\@abstract\@empty\else{\small\@abstract\par\vskip20\p@}\fi
  }

\DeclareRobustCommand\em
        {\@nomath\em \ifdim \fontdimen\@ne\font >\z@
                       \upshape \else \slshape \fi}

\def\@begintheorem#1#2{\sl \trivlist \item[\hskip \labelsep{\bf #1\ #2}]}
\def\@opargbegintheorem#1#2#3{\sl \trivlist
     \item[\hskip \labelsep{\bf #1\ #2\ (#3)}]}

  \def\@arabic#1{\number #1} 

\long\def\@makecaption#1#2{
    \vskip\abovecaptionskip
    \sbox\@tempboxa{{\small {\bf #1}: #2}}%
    \ifdim\wd\@tempboxa>\hsize
        {\small {\bf #1}: #2\par}
    \else
       \global\@minipagefalse
       \hbox to\hsize{\hfil\box\@tempboxa\hfil}
    \fi
    \vskip \belowcaptionskip}

\def\figstrut#1{\hbox to\linewidth{\vrule height#1\hfill}}

\makeatother

\newtheorem{defn}{Definition} [chapter]

 \shtitle{What is a Systems Approach?}  
 \title{What is a Systems Approach?}
 \author{Alex Ryan\affil{alex.ryan@gmail.com}}
 \abstract{What is a systems approach? The first step towards answering this question is an understanding of the history of the systems movement, which includes a survey of contemporary systems discourse. In particular, I examine how systems researchers differentiated their contribution from mechanistic science -- but also from holistic doctrines; and identify the similarities and sharpest differences between complex systems and other systems approaches. Having set the scene, the second step involves developing a definition of `system' consistent with the spirit of the systems approach.}

\begin{document}           
\maketitle

\section{Introduction}
\begin{quote}
\emph{If someone were to analyse current notions and fashionable catchwords, they would find `systems' high on the list. The concept has pervaded all fields of science and penetrated into popular thinking, jargon and mass media.}
\begin{flushright}
Ludwig von Bertalanffy, 1968
\end{flushright}
\end{quote}

Since von Bertalanffy's \cite{von50} theory of open systems introduced the idea of a General Systems Theory (GST) which rose to prominence in the mid twentieth century, the field of systems research has become ever more fashionable, and the closely knit couple of GST and cybernetics have given birth to a large family of systems approaches, including complex systems \cite{Gel94, Kau93, Bar97}, nonlinear dynamical systems \cite{GH83}, synergetics \cite{Hak83}, systems engineering \cite{Hal62}, systems analysis \cite{Dig89}, systems dynamics \cite{For61}, soft systems methodology \cite{Che81}, second order cybernetics \cite{von79, MV84}, purposeful systems \cite{Ack67}, critical systems thinking \cite{Ulr83, Jac85, Jac91}, total systems intervention \cite{FJ91}, and systemic therapy \cite{BJHW56}. As well as adding systems concepts to the tool sets of all fields of science, the systems approach has opened up new areas within disciplines, such as systems biology \cite{Kit02}, and created new hybrid disciplines at the interface between traditional disciplines, such as sociobiology \cite{Wil75}. Many of these systems approaches are introduced in Midgley's \cite{Mid03} epic four volume collection on systems thinking. Diversity is clearly a major strength of the systems approach, but this also makes it difficult to characterise. Consider the following typical definition of a system:
\begin{defn}[System (5)]
A system is a set of entities with relations between them \cite{Lan95}.
\end{defn} \label{defSys1}
By this definition, the converse of a system is a set of entities with no relations, not even logical ones, between them: a heap\footnote{Aristotle \cite{Ari04} used the concept of a heap to refer to matter without form.}. But heaps cannot exist physically, they are only an idealisation, since logical relations can always be established between a set of entities. Alternatively, it could be argued that an indivisible `atomic' element does not meet the definition (provided trivial sets and relations are excluded). Thus a system is never undifferentiated. Nevertheless, under this definition, just about everything bigger than an electron is a system, which makes `system' a vacuous container concept until it is further qualified. For example, the following systems engineering definition only considers physical systems in functional relationships:
\begin{defn}[System (6)]
A system is a bounded region in space-time, in which the component parts are associated in functional relationships \cite{BF98}.
\end{defn}
While there have been many attempts consistent with this definition to make systems the Furniture of the World\footnote{That is, give systems an ontological status or ``real'' existence.} \cite[vol. III]{Bun74}, claims that the world is systemic -- usually synonymous with na\"{\i}ve realist and `hard' systems research -- are problematic. They ignore the fact that systems thinking involves its own set of simplifying assumptions, modelling choices and reductions. They also ignore the insights of second order cybernetics \cite{von79}, that the observer (and by extension the systems theorist) should also be considered as a system. However, there is no need for systems approaches to make the bold claim that the world is ``made up of'' systems. Knowledge that is obtained using a systems approach makes more sense when it is seen as one perspective for thinking about the world, rather than an objective property of bounded regions of space-time. Consequently, a definition of  system should define how the real world is idealised, represented and acted upon when viewed as a system. A definition of system will not tell us how to discern when a part of the world ``is a system'', but it can shed light on when it may be appropriate to utilise a systems approach.

It is also important to stress that the opposite extreme of a relativist and purely subjective account of systems, where all views are equally valid, is not appropriate. In this paper, I will take a moderate approach that considers the insights from both hard and soft systems approaches.
The utility of a systems perspective is the ability to conduct analysis without reducing the study of a system to the study of its parts in isolation. Knowledge obtained from the systems perspective can be as ``objective'' as knowledge obtained from a reductionist scientific perspective. In this way, a systems approach contributes valuable insight capable of complementing the traditional analytic method.

This paper begins by describing the scientific climate prior to the systems movement. Next, the history of systems thinking is interpreted. There are many different stories that could be told to make sense of the development of the systems approach\footnote{Alternatives include Checkland \cite{Che81} and Matthews \cite{Mat04}, or see Lilienfeld \cite{Lil78} for a particularly spiteful critique.}. This particular history is selectively biased towards the design of complex systems. It emphasises the key concepts that differentiate between members of the systems family. The systems approaches I survey are general systems theory, cybernetics, systems analysis, systems engineering, soft systems and complex systems. Finally, I develop a definition of `system' as a concise summary of the systems approach.

\section{Science Before Systems} \label{secScience}
If the Renaissance was a period of re-discovery of classical Greek science, then the subsequent period of the Enlightenment\footnote{The Enlightenment usually refers to the period between the signing of the peace accord at Westphalia in 1648, which brought stability to Western Europe, and the publication of Kant's \cite{Kan81} \emph{Critique of Pure Reason} in 1781, which mounted a sceptical challenge to the Enlightenment philosophy.} produced the scientific revolution that provided a foundation for the Modern worldview \cite{Mat04}. Descartes, the `father of modern philosophy', played a pivotal role in the self-understanding of Enlightenment science. In an attempt to demarcate between knowledge derived from science/philosophy and superstition, Descartes described a scientific method, the adherence to which he hoped could provide privileged access to truth. Descartes' \cite{Des60} method contained four precepts:
\begin{quote}
The first was never to accept anything for true which I did not clearly know
to be such; that is to say, carefully to avoid precipitancy and prejudice,
and to comprise nothing more in my judgement than what was presented to
my mind so clearly and distinctly as to exclude all ground of doubt.

The second, to divide each of the difficulties under examination into as many
parts as possible, and as might be necessary for its adequate solution.

The third, to conduct my thoughts in such order that, by commencing with
objects the simplest and easiest to know, I might ascend by little and
little, and, as it were, step by step, to the knowledge of the more complex;
assigning in thought a certain order even to those objects which in their
own nature do not stand in a relation of antecedence and sequence.

And the last, in every case to make enumerations so complete, and reviews
so general, that I might be assured that nothing was omitted.
\end{quote}
The first rule of sceptical inquiry, and the fourth rule of broad and complete analysis, describe the dominant approach to Modern Western philosophy\footnote{The first rule is hardly a contentious principle in philosophy, but the fourth rule is equally important. For instance, Broad \cite[p. 12]{Bro25} claims that ``the greatest mistake in philosophy is that of over-simplifying the facts to be explained.''}. Meanwhile, the second rule of analytic reduction, and the third rule of understanding the simplest objects and phenomena first, became influential principles of Modern science, which would eventually differentiate itself from philosophy. Together, these two principles provided the view of scientific explanation as decomposing the problem into simple parts to be considered individually, which could then be re-assembled to yield an understanding of the integrated whole. It is this that I will refer to as the Cartesian analytic method.


While Descartes had articulated a philosophy for the scientific method, Newton's \cite{New87} breakthroughs in gravitation and the laws of motion showed the immense power of simple, precise mathematical idealisations to unify and at the same time quantitatively predict the behaviour of diverse phenomena. Newton's Law of Universal Gravitation
\begin{equation} \label{eqnGravity}
F= \frac{Gm_1m_2}{r^2}
\end{equation}
gives the force of attraction $F$ between two objects idealised as point masses $m_1$ and $m_2$, where $r$ is the distance between them and $G$ is a universal constant. The law describes the mechanics of gravity by a universal fixed rule, which captures the first-order effects that dominate the dynamics of most macroscopic inanimate bodies. In particular, our solar system is dominated in terms of mass by the Sun, and the inverse square relationship of gravitational force to distance means that other solar systems exert insignificant forces on our own. This allowed Newton to ignore the great majority of objects in his calculations of planetary motion -- of the $10^5$ objects in our solar system, Newton only had to consider 10 \cite[p. 13]{Wei01}. Further, by Equation (\ref{eqnGravity}), the force between $m_1$ and $m_2$ is independent of any other mass $m_i$. This allows each pair of interactions to be considered independently and then summed: superpositionality holds for the Law of Universal Gravitation, allowing tremendous simplification. Not only did Newton's mechanics provide an alternative cosmology that ended the dominance of Aristotle's worldview, it unified terrestrial and celestial dynamics. Once the heavens and Earth were seen to be governed by the same laws, it became conceivable that the mathematics that accounted for the motion of planets could also account for life on Earth.


Newton's mechanics played a primary role in the $19^{\mathrm{th}}$ century physics worldview of a deterministic, mechanistic Universe, which von Bertalanffy \cite{von56} characterised as the view that ``the aimless play of the atoms, governed by the inexorable laws of mechanical causality, produced all phenomena in the world, inanimate, living, and mental.'' Polanyi \cite[pp. 6-9]{Pol62} traces the origins of mechanism to Galileo's synthesis of a Pythagorean belief that the book of nature is written in geometrical characters, and Democritus' principle: ``By convention coloured, by convention sweet, by convention bitter; in reality only atoms and the void.'' Even though Newton did not personally emphasise this in his philosophy, and many classical mechanists were troubled by the implications of action at a distance, Newton's followers unequivocally interpreted Universal Gravitation as the mechanistic ideal. Laplace, who made important extensions to Newton's physics, provided one of the more famous articulations of mechanism. It became known as Laplace's demon, after the following passage of \emph{Essai philosophique sur les probabilit\`{e}s} \cite{Lap95}.
\begin{quote}
We may regard the present state of the universe as the effect of its past and the cause of its future. An intellect which at a certain moment would know all forces that set nature in motion, and all positions of all items of which nature is composed, if this intellect were also vast enough to submit these data to analysis, it would embrace in a single formula the movements of the greatest bodies of the universe and those of the tiniest atom; for such an intellect nothing would be uncertain and the future just like the past would be present before its eyes.
\end{quote}
For all of its achievements, the Cartesian analytic method in conjunction with mechanism cast a shadow over subsequent Modern science, by setting the laws of theoretical physics as an ideal to which the less exact sciences should aspire, and to which they might eventually be reduced. In this view, only the mechanical properties of things were primary, and the properties studied in other sciences were derivative or secondary \cite{Pol62}.

However, there were many areas of science that resisted the mechanical worldview. The debate between the mechanists and the vitalists on the distinction between inanimate matter and life dominated biology for three centuries. Roughly, the mechanists claimed that life was nothing-but chemistry and physics. Meanwhile, the vitalists countered that life was irreducible to the laws of physics, and the additional vital aspect of living organisms was capable of violating the laws of physics, or at least under-constrained by physics: life was in part self-determining. This was resolved only in the 1920s when the emergentists \cite{Ale20, Ale20a, Bro25, Llo23, Llo26, Llo33} advanced a middle path that acknowledged the need for adherence to the laws of physics, while at the same time denying that all phenomena could be eliminatively reduced to physics. In retrospect, emergentism marked a pivotal advance towards systems thinking.


Checkland \cite{Che81} identified three further classes of problems that persistently failed to be conquered by the Cartesian analytic method of science. They are problems of complexity, problems of social science, and problems of management. According to Checkland,
the complexity that arises when densely interacting variables give rise to `emergent phenomena' poses a serious problem, one that to date reductionist thinking and the experimental method has not been able to overcome.

Secondly, the problems of social science are not just densely interacting, but they contain individual human beings with potentially unique private knowledge and interpretations that condition their response, limiting the precision and general applicability of `laws' governing social behaviour. In addition, social phenomena can always be interpreted from many more possible perspectives than are required for natural science.

Thirdly, Checkland cites management -- the process of taking decisions in social systems -- as problematic for analytic science. Operations Research (OR), the scientific approach to management, arose in support of military operational decisions during the second world war, and was institutionalised and applied widely in industry after the war. However, Checkland claims that OR's ability to solve problems with particular general forms has been of little use to real life problems, where the details that make a problem unique predominates any insight that knowledge of the general form provides. Even though Checkland does not substantiate this assertion, it is reinforced by observations from within the OR community that OR is being increasingly relegated to the tactical rather than strategic decision-making arena \cite{Eil80}:
\begin{quote}
the uniqueness of the OR approach is not seen as indispensable, its methodology is challenged, it is regarded as a narrow specialist discipline, a suitable sanctuary for mathematicians, its involvement in implementation is tenuous and its general impact somewhat limited.
\end{quote}

Whilst it would be a gross oversimplification to say that science prior to the mid twentieth century was non-systemic, it was not until this period that a self-consciousness of what it meant to be systemic arose. It is in the context of a dominant analytic approach to science, and a growing awareness of its limitations, that the systems movement sought to provide an alternative approach. The common usage of `system' within analytic science prior to the systems movement meant simply the object or objects of interest. However, this usage does not attribute any properties to a system -- it acts as a container, merely a convenient label for the collection of interacting objects under investigation. In contrast, the systems movement began by articulating a more substantial conception of system.

\section{Enter the System} \label{secEnterSys}
In the West, von Bertalanffy's \cite{von50} seminal paper on open systems, published in 1950, is usually attributed as seeding the rise of the systems movement, although he had published on a systems approach to biology since 1928 \cite{von33}. The purpose of the paper was to rigorously account for the key distinction between the organismic systems of biology, compared with the  closed systems of conventional physics. In making this distinction, von Bertalanffy wished to scientifically account for apparent paradoxes in the characteristics of living systems, when considered in relation to the physics of closed systems. For example, the second law of thermodynamics states the inevitability of increasing entropy and the loss of order, and yet the evolution and development of biological systems can exhibit sustained increases in order. By providing empirically supported scientific explanations of phenomena such as equifinality (when a system's dynamics are not state-determined) and anamorphosis (the tendency towards increasing complication), von Bertalanffy simultaneously extinguished vitalism and firmly  established the roots of systems theory in the natural sciences.

Von Bertalanffy's emphasis on flows of matter (and later energy and information) into and out of an open system brought attention to the environment of the open system. This adds something that Definition \ref{defSys1} does not contain: a system is more than just a set of components and their relationships -- it is a complex whole that affects and is affected by its environment. Further, a system has a boundary that prevents it from becoming mixed with its environment. The implication of the environment is that a system must always be understood in context.

\begin{figure}[!htb]
\begin{center}
\epsfig{file=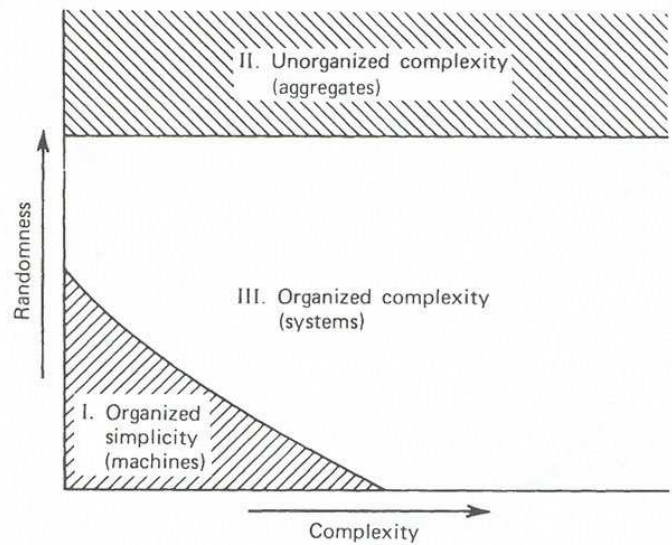,width=0.75\textwidth}
\end{center}
\caption{Types of systems, after Weinberg \cite{Wei01}.}
\label{figoc}
\end{figure}

The domain of systems theory does not cover all systems -- it was never intended as a theory of everything. In 1948, Weaver \cite{Wea48} noted that between mechanics and statistical mechanics, there was an absence of mathematical techniques for the treatment of systems with medium numbers of components, whose behaviour could be much more varied than either simple machines or very large ensembles. Weaver gave a name to the systems in this gap, saying that ``one can refer to this group of problems as those of \emph{organized complexity}''. Weaver illustrated the domain of organised complexity graphically. A slightly elaborated version of this graphic by Weinberg \cite{Wei01} is reproduced in Figure \ref{figoc}.
In Weaver's view, the entities that systems theory studies are substantial (physical) systems. Occasionally, confusion has arisen as to whether the domain of systems theory extends to the study of abstract (conceptual) systems. (Ashby \cite[p. 260]{Ash62} notes this general uncertainty within GST.) This confusion is exacerbated because systems theories of both persuasions have been developed, often without explicitly addressing their domain of applicability. However, it is fair to say that both substantial and abstract systems have been legitimate subjects of systems enquiry.

Checkland \cite[pp. 74-92]{Che81} suggested that systems thinking was founded on two pairs of ideas:
\begin{enumerate}
\item Emergence and hierarchy, originating in organismic biology and generalised in GST; and
\item Communication and control, originating in communication engineering and generalised in cybernetics.
\end{enumerate}
To these pairs of ideas should be added the requirement for a systems approach to be broader than any conventional discipline: a systems approach is fundamentally interdisciplinary. In retrospect, a sophisticated theory of systems that preceded both GST and cybernetics had been developed by Bogdanov \cite{Dud96} in Russia and published sometime between 1910-1913, but for political reasons his Tektology was not widely known until after the Cold War, by which time the systems movement was already well established in the West, and synonymous with GST and cybernetics.

Another precursor to a recognised systems movement (by about a decade) was Angyal's \cite{Ang41} theory of systems, developed in the context of psychology. In particular, Angyal distinguished between relations and systems:
\begin{quote}
A relation requires two and only two members (relata) between which the relation is established. A complex relation can always be analysed into pairs of relata, while the system cannot be thus analysed.
\end{quote}
This distinction leads to the difference between aggregates, in which the parts are added, and systems, in which the parts are arranged. Note that Newtonian mechanics only describes the behaviour of aggregates. Angyal also realised the importance of the manifold that the system is embedded in, saying that a system is ``a distribution of the members in a dimensional domain.'' Hence, a system is more than a set of interrelated  components: the relations must be \emph{organised}. Spatial arrangements are an important determinant of systemic properties. This also implies that many combinations of relations will not be possible for any particular system, since the relations must conform to the system's organisation. Thus organisation imposes order on a system, which can be thought of as a constraint on its dynamics.

Because the systems approach stood in contrast to mechanism (and also the related `isms' of atomism and individualism), some began to associate systems with holism. The systems philosopher Bunge \cite{Bun77a, Bun00} recognised this conflation and sought to distinguish systemism from holism. According to Bunge \cite{Bun77a},
\begin{quote}
a holistic philosophy is one that not only accepts the thesis that a whole is more than a mere aggregation of its parts: holism maintains also that wholes must be taken at face value, understood by themselves, not through analysis.
\end{quote}
Because the holistic approach rejects the possibility of analysis, it relies upon the method of intuition, not rational explanation or empirical experiment. While the systems approach recognises the existence of emergent properties, it nevertheless seeks to explain them in terms of how their constituent parts are organised. Where holism is satisfied with a non-rational apprehension of un-analysed wholes, systemism aims to demystify emergent properties by providing scientific understanding that utilises analysis as well as synthesis. Therefore, it is equally important that the systems approach be distinguished from holism as from mechanism.

Another important refinement to the philosophical characterisation of the systems approach was provided by Churchman \cite{Chu68}, in the form of the following question:
\begin{quote}
How can we design improvement in large systems without understanding the whole system, and if
the answer is that we cannot, how is it possible to understand the whole system?
\end{quote}
Ulrich \cite{Ulr02}, a student and self-proclaimed disciple of Churchman, realised that Churchman's question captured ``the real challenge posed by the systems idea: its message is not that in order to be rational we need to be omniscient but, rather, that we must learn to deal critically with the fact that we never are.'' Thus, in a very deep sense, the systems approach is tied to understanding the limits of representations.

\subsection{General systems theory} \label{secGST}
Von Bertalanffy together with Rapoport and Boulding formed the Society for General System Theory in 1954, which organised the systems community and provided a yearbook from 1956 dedicated to systems research. Along with the closely related field of cybernetics, GST helped to define the core principles of the systems approach.

For von Bertalanffy, the main propositions of GST (adapted from \cite{von56}) were:
\begin{enumerate}
\item Isomorphisms between the mathematical structures in different scientific disciplines could integrate and unify the sciences;
\item Open systems require consideration of the flow of energy, matter and information between the system and its environment;
\item Within open systems, the same final system state may be reached from different initial conditions and by different paths -- open systems exhibit equifinality;
\item Teleological behaviour directed towards a final state or goal is a legitimate phenomenon for systemic scientific inquiry;
\item A scientific theory of organisation is required, to account for wholeness, growth, differentiation, hierarchical order, dominance, control and competition; and
\item GST could provide a basis for the education of scientific generalists.
\end{enumerate}
Von Bertalanffy's principle concern was to provide an alternative foundation for unifying science, which he proposed in reaction to the reductionist mechanistic worldview. In particular, he rejected the crude additive machine theories of organic behaviour, which treated wholes as nothing more than linear aggregates of their components. It is notable that von Bertalanffy \cite{von56, von69} really only mentions emergence and hierarchy in passing. GST adopted almost unchanged the theory of biological emergentism developed in the 1920s, while the task of developing hierarchy within GST was taken up by other authors. In fact, I would suggest that rather than a theory of general systems, GST resulted in the more modest contribution of several theories of hierarchies. The remainder of this section on GST will consider the twin concepts of emergence and hierarchy.

Emergence as ``the whole is more than the sum of the parts'' justified the need to understand systems in addition to understanding their parts. By understanding emergent properties, the general systems theorists felt that they could offer insights that the reductionist / mechanist agenda ignored, because mechanism could not account for the non-additive relationships between components. The ability to explain emergent properties is a prerequisite for a general theory of organisation.

The hierarchical nature of systems is a logical consequence of the way system is defined in terms of its constituent parts, since the parts may also meet the definition of system. This was noted by many authors, but Boulding and Simon made particularly influential extensions to this corollary. Boulding's \cite{Bou56} \emph{Skeleton of Science} presented a hierarchical view of systems, ``roughly corresponding to the complexity of the `individuals' of the various empirical fields.'' Where von Bertalanffy \cite{von50} had focussed on the relationship of physics and chemistry with biology, Boulding generalised this into a hierarchy with nine levels, which began with physics and extended through biology, psychology, sociology and metaphysics. Following Matthews \cite[p. 201]{Mat04}, Boulding's hierarchy is summarised in tabular form in Table \ref{tabBoulding}.

Boulding saw these levels as different layers of theoretical discourse. That is, the layers were not part of nature itself, but were a way of organising the theoretical concepts and empirical data collected in the various sciences (although it has not always been presented this way by other authors). When Boulding later aligned specific disciplines with the levels, the intent was to highlight that, for example, any analysis of level 8 sociocultural systems using only the methods of level 1 and 2 structures and mechanics would be necessarily incomplete. Boulding's hierarchy has been extended by Checkland \cite{Che81} and critiqued by numerous authors including Capra \cite{Cap96}. Despite its shortcomings, Boulding's hierarchy was an influential representation of the relations between the sciences. It also reveals a view that many early systems theorists shared: that their role was to be one step removed from empirical science, in order to recognise the broader patterns occurring across science. Others, such as Bunge \cite{Bun73} and Rapoport \cite{Rap76}, later developed more explicit arguments that the systems approach lies midway between the scientific and philosophical approaches.

\setlength\extrarowheight{4pt} 
\begin{threeparttable}
\caption{Boulding's hierarchy of systems complexity.} \label{tabBoulding}
\begin{tabular}{|>{\raggedright}p{0.18 \textwidth}|>{\raggedright}p{0.26 \textwidth}
|>{\raggedright}p{0.22 \textwidth}|>{\raggedright}p{0.23 \textwidth}|} \hline
\textbf{Level} & \textbf{Characteristics} & \textbf{Examples} & \textbf{Relevant Discipline} \tabularnewline \hline
1. Structure & Static & Crystals & Any \tabularnewline \hline
2. Clock-works & Pre-determined motion & Machines, the solar system & Physics, Chemistry \tabularnewline \hline
3. Control mechanisms & Closed-loop control & Thermostats, mechanisms in organisms & Cybernetics, Control Theory \tabularnewline \hline
4. Open systems & Structurally self maintaining & Flames, biological cells & Information Theory, Biology (metabolism) \tabularnewline \hline
5. Lower organisms & Organised whole with functional parts, growth, reproduction & Plants & Botany \tabularnewline \hline
6. Animals & A brain to guide total behaviour, ability to learn & Birds and Beasts & Zoology \tabularnewline \hline
7. Humans & Self-consciousness, knowledge symbolic language & Humans & Psychology, Human
Biology \tabularnewline \hline
8. Socio-cultural systems & Roles, communication, transmission of values & Families, clubs,
organisations, nations & Sociology, Anthropology \tabularnewline \hline
9. Transcendental systems & Inescapable unknowables & God & Metaphysics, Theology \tabularnewline \hline
\end{tabular}
\end{threeparttable}

\vspace{\baselineskip}

Simon's \cite{Sim62} classic paper on \emph{The Architecture of Complexity} sought to explain some features of naturally arising hierarchies, in contrast to Boulding's attempt to relate the description of nature within different disciplines of science. Simon proposed that systems with a large number of non-simply interacting parts frequently take the form of hierarchy, defined as
\begin{quote}
a system that is composed of interrelated sub-systems, each of the latter being, in turn, hierarchic in structure until we reach some lowest level of elementary subsystem.
\end{quote}
Simon is describing a nested hierarchy, which was later generalised in the highly abstract GST dialect of hierarchy theory \cite{All06} that this paper helped to initiate. There are two central insights in \cite{Sim62}. Firstly, what Simon saw as the ubiquity of hierarchies in natural complex systems is explained by a simple probability argument: the time it takes for the evolution of a complex form depends critically on the number of potential intermediate stable forms, as these building blocks to a large degree insulate the process of system assembly against the effects of environmental interference. Given that a hierarchy of building blocks can be assembled orders of magnitude faster than a non-hierarchical assembly process, among complex forms, hierarchies are the ones that have the time to evolve.

Secondly, Simon realised that the interactions at each level of the hierarchy are often of different orders of magnitude, and commonly the interactions are strongest and most frequent at the lowest level in the hierarchy. When these conditions hold, the hierarchy is \emph{nearly decomposable}, which simplifies the analysis of a complex system in several ways. Near decomposability implies that subparts belonging to different parts only interact in aggregate, meaning individual interactions can be ignored: the levels are screened off from each other by rate differences. This is why in modern hierarchy theory, three levels of a hierarchy, one up and one down, are generally considered sufficient for analysis of the focal level. Also, nearly decomposable hierarchies are highly compressible: they have significant redundancy, meaning that a relatively small amount of information may adequately describe an apparently complex system.

\subsection{Cybernetics} \label{secCybernetics}
At about the same time that GST was attempting a general mathematics of organisation, the new field of cybernetics was embarking on a similar quest to uncover the general mathematics of machines. According to Wiener \cite{Wie48}, who suggested the name for the field,
\begin{quote}
cybernetics attempts to find the common elements in the functioning of automatic machines and of the human nervous system, and to develop a theory which will cover the entire field of control and communication in machines and in living organisms.
\end{quote}
In \cite{Wie48}, Wiener described the initial results from the cybernetic community, whose impressive interdisciplinary role-call included Rosenblueth, von Neumann, McCulloch, Pitts, Lorente de N\'{o}, Lewin, Bateson, Mead and Morgenstern. With the exception of Morgenstern, they were all members of the core group of ten conferences on cybernetics held between 1946 and 1953, sponsored by the Josiah Macy, Jr. Foundation and chaired by McCulloch. Wiener identified the principle of feedback as a central step towards the study of the nervous system as an integrated whole, outlining the role of negative feedback in control, and positive feedback in producing instability. The principle of feedback distinguishes cybernetic systems by their circular organisation, where activity flows from effectors acting on the environment, to sensors detecting their outcome, which then act upon the effectors, thereby closing the feedback loop. The other major distinction that Wiener drew was between communication engineering and power engineering as a basis for understanding mechanical systems. He identified a change in emphasis in the former from the economy of energy to the accurate reproduction of a signal. The implications of this distinction are that understanding the behaviour of a control system depends not on the flow of energy so much as the flow of \emph{information}. Fortuitously, at the same time Shannon \cite{Sha48} was developing a quantitative definition of information as the average reduction in uncertainty, which became one of the dominant frameworks within cybernetics, and more generally within systems theory.

Although Wiener's \cite{Wie48} analogies between calculating devices and the human brain now appear dated, a number of the big systems ideas arose from Wiener's cybernetic collaborators in attempting to understand the behaviour of ``all possible machines''. They include von Neumann's self-reproducing automata \cite{von66}, McCulloch and Pitts' \cite{MP43} model of the neuron that forms the basis for artificial neural networks, Maturana's \cite{MV84} autopoesis, von Foerster's \cite{von79} second order cybernetics and Bateson's \cite{Bat72} ecology of mind, each exploring different implications of circularity. 

Outside the US, the British psychiatrist Ashby had published on what became known as cybernetics from 1940, inventing the homeostat -- a machine that adapted towards self-regulation -- and writing the 1956 classic text \emph{An Introduction to Cybernetics}. Ashby's ideas have had a profound influence on complex systems theory, pre-empting and shaping the complex systems approach, and still providing insights on contemporary concerns such as self-organisation, adaptation and control. Perhaps Ashby's greatest contributions were the connections he developed between information theory and systems theory. In \cite[p. 3]{Ash56}, Ashby observed
\begin{quote}
cybernetics typically treats any given, particular, machine by asking not ``what individual act will it produce here and now?'' but ``what are \emph{all} the possible behaviours that it can produce?''

It is in this way that information theory comes to play an essential part in the subject; for information theory is characterised essentially by its dealing always with a \emph{set} of possibilities \ldots
\end{quote}


Ashby's most famous result, introduced in this book, applies information theory to machine regulation to yield the law\footnote{The law of requisite variety is actually a mathematical theorem, not a physical law.} of requisite variety. The law provides a quantitative measure of the amount of variety a regulator $R$ must absorb in order to produce goal-directed behaviour. Ashby realised that the purpose of a system could be re-cast as requiring essential system variables, $E$, to be maintained within certain limits. He argued that natural selection has eliminated those organisms that are not able to regulate the flow of environmental variability. The remaining species of organisms all employ mechanisms that actively resist environmental disturbances, $D$, that would push the essential variables outside their limits and result in death.

The law is simple to introduce. If different disturbances can be countered by the same response by the regulator, they are not counted\footnote{This way of counting disturbances does a lot of the work in the law of requisite variety, and it is the key assumption that must be satisfied for real world applications of Ashby's law to be valid.} as distinct members of $D$. If the set of possible outcomes, $O$, of the disturbance followed by regulation (denoted by the transformation matrix $T$) is within the acceptable limits of $E$, it is compatible with the organism's continued existence (denoted by the set $\eta$). In this case, the regulator is considered successful. Let the variety, measured as a logarithm of the number of possibilities, of $D, R$ and $O$, be $V_D, V_R$ and $V_O$ respectively. Then the law of requisite variety is simply
\begin{equation} \label{eqnAshbyLaw}
V_O \le V_D - V_R.
\end{equation}
Intuitively, goal-directed behaviour demands that variety in the outcome remain below some bound. In order to reduce $V_O$, $V_R$ must be increased: only variety in the regulator can destroy variety created by the environmental disturbance. Another\footnote{Thanks to Scott Page for this suggestion.} way to understand the law of requisite variety is to ask what if $R>D$?. Then by Equation \ref{eqnAshbyLaw}, $V_O < 0 \Rightarrow O < 1$. This means that the disturbances are perfectly countered by the regulator, and so there is no variety in the outcome of the process.

\begin{figure}[!htb]
\begin{center}
\includegraphics[scale=0.6]{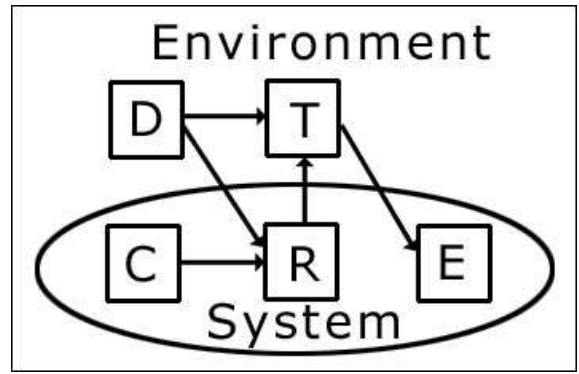}
\end{center}
\caption{Control and regulation in Ashby's law of requisite variety.}
\label{figAshby}
\end{figure}

The law of requisite variety sheds light on the nature of control. Figure \ref{figAshby}, adapted from \cite{Ash56}, shows the causal influences between the variables introduced above and a new control variable $C$. In this model, $C$ decides the desired outcome or sequence of outcomes, which $R$ must obey. Roughly, $C$, $R$ and $E$ constitute the organism or the `system', which is open to energy and information but closed with respect to control. $R$ takes information from the independent sources, $C$ and $D$, and attempts to produce an action that achieves $C$'s objective in spite of the external disruption $D$. The `environment' $T$ transforms the actions of $D$ and $R$, which influences the value of the essential variables $E$. Therefore, control over $E$ \emph{necessarily requires} regulation. This can be interpreted as the communication of a message from $C$ to $E$ via the compound channel $T$, while transmitting nothing from $D$. This is not just a superficial analogy to information theory: the law of requisite variety is a restatement of Shannon's \cite{Sha48} tenth theorem on channel coding. This reveals the deep connections between goal-directed behaviour, control, regulation and communication.

Another connection between information theory and systems is evident in Ashby's definition of complexity as the quantity of (Shannon) information required to describe a system \cite{Ash73}, although Ashby was not the first to propose this definition.

Where GST predominantly analysed structure to understand organisation, cybernetics developed a complementary approach for analysing dynamical \emph{behaviour}, independently of how the system was internally structured. The two are complementary because structure can be viewed as a constraint on the system's dynamics, while the dynamics are responsible for the formation of structure. The behaviourist approach to modelling systems was formalised within cybernetics by generalising the electrical engineer's `black box'. The black box approach assumes that the internal structure of a system is hidden, and so knowledge of the system must be obtained by systematically varying inputs to the system and observing the corresponding outputs. The same approach to behavioural psychology is described as stimulus-response (as in \cite{Wat19}), and both terminologies were used interchangeably in cybernetics.

Although GST and cybernetics were strong allies in advancing the systems movement, there were some important differentiators. The research agenda for cybernetics was focussed on machines, while GST was much broader. The commitment of cybernetics to behaviourist mechanical explanations outlined in the previous paragraph stands in contrast to the position within GST that mechanics provides an incomplete account of open systems. Thus in Boulding's hierarchy, systems at or above level four require more than just an understanding of information and control as provided by cybernetics at level three. Also, in cybernetics emergence was typically dismissed as incompleteness of the observer's knowledge of the parts and their couplings (see for example \cite[pp. 110-112]{Ash56}). The only time cybernetics acknowledged genuine emergence was as a result of self-reference or circularity. Unlike GST, cybernetics embraced mechanistic explanations, but sought to augment the physicists' view of causes preceding effects with circular causation and goal-directed behaviour.

\subsection{Systems analysis}
Systems analysis was an extension of Operations Research (OR) to broader concerns. Morse and Kimball's \cite{MK51} book set the standard for OR, including its definition.
\begin{quote}
Operations research is a scientific method of providing executive departments with a quantitative basis for decisions regarding the operations under their control.
\end{quote}
Although slightly older, OR had some common ground with the new systems approaches. In particular, OR too saw itself as an interdisciplinary approach based on scientific methods. The way in which OR provided a scientific basis for decisions was to use mathematical models. Also, OR often developed integrated solutions, rather than considering individual components in isolation. Network flow, queueing theory and transportation problems are typical examples of this, but also serve to highlight some key differences. Traditional OR addressed problems where the objectives were precisely given, and the components were fixed \cite{FS93}, but their configuration could change according to the value of the control parameter $\mathbf{x}$. The operations researcher would structure the problem by framing it in such a way that the performance of alternative configurations could be compared against the same objective. This focus on increasing measurable efficiency meant that OR was traditionally interested in only one organisation: the configuration that gives the global minimiser of the objective function $f(\mathbf{x})$. As OR developed increasingly sophisticated mathematical techniques, it became more theoretical and less interdisciplinary, which led to a number of issues that have already been mentioned in Section \ref{secScience}.

In the newly established RAND (Research ANd Development) Corporation, the mathematician Paxson (whose work has not been declassified) was more interested in decisions affecting the next generation of military equipment, than configuring a fixed set of platforms already in operational service. Early applications of OR to what Paxson called ``systems analysis'' were criticised because they did not adequately consider costs \cite{Dig89}. Consequently, systems analysis became a collaborative venture between mathematicians/engineers and economists. Successful high profile projects in air defence and the basing of bombers added ``Red Team'' game-theoretical approaches to the systems analysis methodology, generated sustained interest in systems analysis within RAND, and gained the attention of senior management. As well as borrowing from and extending OR, systems analysis inherited the black boxes and feedback loops from cybernetics in order to construct block diagrams and flow graphs. A subsequent technique for performing systems analysis -- Forrester's \cite{For61} system dynamics\footnote{System dynamics represented the world as a system of stocks and flows, from which the behaviour of feedback loops could be deduced.} -- made more explicit use of these cybernetic concepts. After a decade where RAND led the formalisation and application of the systems analysis methodology, in 1961 the Kennedy Administration brought both RAND people and systems analysis techniques into government to provide a quantitative basis for broad decision-making problems \cite{Dig89}. McNamara's use of systems analysis within the Department of Defense and NATO was highly controversial and widely criticised. Nevertheless, it is notable as the systems approach which has had the greatest impact on society, due to the scale and nature of the decisions that it justified. In particular, the Vietnam war was planned with considerable input from systems analysis techniques combining operations research with cost analysis.

Aside from a rudimentary use of feedback, systems analysis was largely independent of developments in systems theory. As Checkland \cite[p. 95]{Che81} nicely described it,
\begin{quote}
on the whole the RAND/OR/management science world has been unaffected by the theoretical development of systems thinking, it has been \emph{systematic} rather than \emph{systemic} in outlook \ldots
\end{quote}
By this, Checkland meant that systems analysts assembled cookbooks of mechanical solutions to classes of recurring problems, rather than developing techniques that addressed emergent properties, interdependencies and environmental influences. In so far as it was applied to the class of simple mechanical systems depicted above in Figure \ref{figoc}, this was perfectly reasonable. However, as the interests of systems analysis broadened from merely technical systems to include many social factors, this omission became more apparent.




\subsection{Systems engineering} \label{secSE}
Systems engineering has a history quite separate to the systems movement. Its closest historical link comes from the application of systems analysis techniques to parts of the systems engineering process. The need for systems engineering arose from problems in the design and implementation of solutions to large scale engineering challenges spanning multiple engineering disciplines. A multidisciplinary team of engineers required a lead engineer whose focus was not the design of individual components, but how they integrated. Consequently, management concerns were as significant as technical challenges for a systems engineer. The Bell Telephone Laboratories and Western Electric Company's design and manufacture of the Nike air defence system, commenced in 1945, is widely cited as one of the first systems engineering projects. 
The surface to air missile defense program integrated ground-based tracking radars, computers and radio controlled anti-aircraft rockets, in order to protect large areas from high altitude bombers. It was novel because unlike conventional anti-aircraft artillery, Nike allowed continuous missile guidance: the radars and computers enabled feedback and control. Bell Labs were the prime contractor for the project, while much of the detailed engineering was undertaken by the major subcontractor, Douglas Aircraft Company. The 1945 Bell Labs report \emph{A Study of an Antiaircraft Guided Missile System} was considered a classic in applied interdisciplinary research due to its depth of insight, scope, and influential role in the systems engineering of Nike \cite{Fag78}.

Following the success of individual systems engineering projects, Bell Labs structured itself around the new systems engineering approach. Bell Labs was organised into three areas: basic research, systems engineering and manufacturing projects \cite{Kel50}. The systems engineering area provided the interface between advances in communications theories and the manufacture of commercial systems. Because of the ``whole system'' perspective within the systems engineering area, it was responsible for prioritising the activation of projects with the greatest potential user benefit, within the technical feasibility of communications theory. The responsibility of the systems engineer was the choice of the ``technical path'' between theory and application in order to create new services; improve the quality of existing services; or lower their cost. Because of its emphasis on user benefit, standards were seen to play a vital role in systems engineering. Standards were used to measure quality of service, which enabled cost benefit analysis of different technical paths.

When framed as the decision-making problem of selecting between different technical paths on the basis of cost effectiveness, systems engineering appears closely related to the problems that RAND were concurrently tackling with systems analysis. Bell Labs was aware of this analogy \cite{Wal66}, and drew on OR and systems analysis techniques. However, the systems engineer typically placed a greater emphasis on technical knowledge than the systems analyst, and correspondingly less emphasis on mathematical models. In contrast to the dynamical models of systems analysis, systems engineers developed architectures to represent system designs.

Outside Bell Labs, Project Apollo was one of the highest profile early successes of the systems engineering approach, which quickly spread from its origins in defence to also become the standard approach to large scale civilian projects. The traditional systems engineering process can be summarised as follows: 1) Customer needs are captured in precise, quantified requirements specifications; 2) System requirements are decomposed into requirements for subsystems, until each subsystem requirements is sufficiently simple; 3) Design synthesis integrates subsystems; and 4) Test and evaluation identifies unintended interactions between subsystems, which may generate additional requirements for some subsystems. If there are unintended consequences (i.e. unplanned emergent properties), the process returns to stage 2, and repeats until the system meets the requirements.


Because systems engineering is applied to the most ambitious engineering projects, it also has a long list of large and public failures \cite{Bar03}. Partly in response to perceived failures, systems engineering has identified ``System of Systems'' problems as a distinct class of problems that are not well suited to traditional centrally managed systems engineering processes. Although there is not a standard System of Systems (SoS) definition, the term SoS usually denotes heterogeneous networks of systems including a majority of Maier's \cite{Mai98} discriminating factors: operational and managerial independence, geographical distribution, and emergent and evolutionary behaviours. The recent popularity of the SoS buzzword in the systems engineering literature has prompted the expansion of systems engineering techniques to include methods that can cope with evolving networks of semi-autonomous systems. This has led many systems engineers to read more widely across the systems literature, and is providing a re-conceptualisation of systems engineering as part of the systems movement, despite its historical independence. This is reflected in the latest INCOSE handbook \cite[p. 2.1]{Has06}, which states ``[t]he systems engineering perspective is based on systems thinking'', which ``recognizes circular causation, where a variable is both the cause and the effect of another and recognizes the primacy of interrelationships and non-linear and organic thinking---a way of thinking where the primacy of the whole is acknowledged''.



\subsection{Soft systems}
The systems approaches I have surveyed so far have each expressed an unbounded enthusiasm for multi-disciplinarity in their missionary papers, and yet all have tended to converge towards a relatively narrow band of `hard' scientific methods, where this is taken to mean strong scepticism towards any theory that cannot be stated exactly in the language of mathematics. C. P. Snow's famous ``Culture Gap'' between literary intellectuals and scientists proclaimed that a gulf between the two cultures prohibited any real communication or shared understanding between them. A related, but perhaps deeper cultural divide was noted a generation earlier by James \cite{Jam96} (as quoted in \cite{Mat04}) in 1909:
\begin{quote}
If you know whether a man is a decided monist or a decided pluralist, you perhaps know
more about the rest of his opinions than if you give him any other name ending in `ist' \dots to believe in the one or in the many, that is the classification with the maximum number of
consequences.
\end{quote}

Even within science, this metaphysical divide can be observed. Another way of approaching the dichotomy is to ask whether the goal of enquiry is objective knowledge or inter-subjective discourse. Hard science, by accepting only mathematically precise arguments, adopts a monistic stance. Consequently, hard science is associated with the search for a \emph{unified rational foundation} for systems universally valid across the sciences. Von Bertalanffy, Wiener, Ashby and Forrester all clearly held this aspiration to varying degrees. In contrast, a pluralist position rejects the possibility of unification, emphasising the incommensurability of frameworks that view reality from different perspectives. This is rejected by the monist, because it is seen to be opening the door to inconsistency, paradox, relativism and irrationality. Within the Macy conferences, issues of subjective experience and \emph{Gestalten} (form perception) were raised, usually by psychoanalysts such as Kubie \cite{Ame06}. The hard scientists sustained vocal criticism that tended to suppress these ideas: in particular, they strongly questioned the scientific status of psychoanalysis.

Von Foerster, an editor of the Macy conference proceedings, later developed second order cybernetics \cite{von79}, which shifted attention from the cybernetics of \emph{observed} systems, to the cybernetics of \emph{observing} systems. In studying the ``cybernetics of cybernetics'', von Foerster applied cybernetic concepts to the observer, which led him to reject the possibility of objective observation. Von Foerster's \cite{von92} variant on James' metaphysical distinction was introduced as
\begin{quote}
a choice between a pair of in principle undecidable questions which are, ``Am I \emph{apart from} the universe?'' Meaning whenever I \emph{look}, I'm looking as if through a peephole upon an unfolding universe; or, ``Am I \emph{part of} the universe?'' Meaning whenever I \emph{act}, I'm changing myself and the universe as well.
\end{quote}
By adopting the latter position, von Foerster was dismissing the search for a unified rational foundation for systems as an ill-posed problem. Even though von Foerster approached second order cybernetics using essentially hard techniques, he played an important role in undermining the doctrine of objectivity that ``[t]he properties of the observer shall not enter the description of his observations'' \cite{von79}. Second order cybernetics provides a scientific basis for the theory-ladenness of observation: the philosophical assertion that all observations contain an inseparable element of theory.

In 1981, Checkland \cite{Che81} published the first sophisticated systems methodology for `soft' problems. Checkland firstly divided systems into five types: natural systems (atomic nuclei to galaxies); designed physical systems (hammers to space rockets); designed abstract systems (mathematics, poetry and philosophy); human activity systems (a man wielding a hammer to the international political system); and transcendental systems (as in Boulding's hierarchy, although these are not considered in detail). Checkland's central thesis was that human activity systems were fundamentally different in kind from natural systems, and consequently required a fundamentally different approach. Checkland \cite[p. 115]{Che81} argued that
\begin{quote}
The difference lies in the fact that such systems could be very different from how they are, whereas natural systems, without human intervention, could not. And the origin of this difference is the special characteristics which distinguish the human being from other natural systems.
\end{quote}
That special characteristic is self-consciousness, which is claimed by Checkland to provide ``irreducible freedom of action'' to humans. However, one should note that this distinction between the human being and other natural systems is much more of a metaphysical assertion than an empirical fact, and it is the most crucial theoretical assertion for Checkland's argument against the use of hard methods in human activity systems.

The fundamentally different approach that Checkland used to investigate human activity systems was `action research'. According to Checkland \cite[p. 152]{Che81},
\begin{quote}
Its core is the idea that the researcher does not remain an observer outside the subject of investigation but becomes a participant in the relevant human group.
\end{quote}
From this, it is clear that action research follows von Foerster's lead in exploring the implications of \emph{acting} as \emph{part of} the universe. This commitment is responsible for the two distinguishing features of the soft systems approach: the way problems are framed, and the way models of the system are built and used.

The difference in problem framing is most clearly visible in Checkland's \cite[p. 316]{Che81} definition of a soft problem:
\begin{defn}[Problem, Soft]
A problem, usually a real-world problem, which cannot be formulated as a search for an efficient means of achieving a defined end; a problem in which ends, goals, purposes are themselves problematic.
\end{defn}
Thus, instead of defining an objective that can be tested at any point in time to see if it has been met, the soft systems approach operates with a comparatively unstructured and fuzzy understanding of the ``problem situation''. Because the problem is framed using only minimal structure, the quantitative methods of OR and systems analysis cannot be brought to bear on its solution. Nor is the aim of soft systems optimal efficiency, instead it seeks feasible, desirable change.

The way models are built is fundamentally pluralist: ``human activity systems can never be described (or `modelled') in a single account which will be either generally acceptable or sufficient'' \cite[p. 191]{Che81}. It is accepted that different people will view the system and its problems differently, and models aim to make explicit the assumptions of a particular view in order to facilitate dialog, rather than building a single foundational representation of the system. Thus, the difference between hard and soft systems approaches is a deep divide: it is James' metaphysical distinction between the belief in the one or in the many. Perhaps the most important point in favour of soft systems methodology is that it has mostly been considered useful in practice for solving real-world problems \cite{Min92a}.


Habermas \cite{Hab72} added a further category in contrast to what I have described as hard and soft approaches, with the development of critical theory. According to Habermas, humans have developed a technical interest in the control and manipulation of the world; a practical interest in communicating to share understanding with other people; and an emancipatory interest in self development and freedom from false ideas. Critical theory aims to reveal systematic distortions resulting from the power structure that affect both hard (technical) and soft (communicative) approaches. It formed the theoretical basis for critical systems theory, which has produced several methodologies \cite{Jac85, Ulr83} for taking practical action in management contexts. However, because the intention of critical theory is emancipation, not design or even intervention, these methods have struggled to resolve this fundamental inconsistency. The main contribution of critical systems theory to date has been to critique both the hard and soft systems approaches, rather than to advance a practical alternative to systems design (see for example \cite{Min92}). Consequently, it will not be considered further in this thesis.


\subsection{Complex systems} \label{secCS}
Complex systems became known as a distinct systems discourse with the establishment of the Santa Fe Institute (SFI). A widely read romantic account of the formation of the SFI and its initial research agenda was published by the science writer Waldrop \cite{Wal92}. Its success set the model for communicating complex systems research to the informed general public through easy to read popular science novels. These novels attempted to capture the excitement of research on the frontier of ``a new kind of science'', while at the same time emphasising its applicability to important global issues, which are inevitably both systemic and complex. The upside of the way complex systems was marketed is that it has grown to become the most active\footnote{This claim is justified by the scale and number of conferences, the number of complex systems centres, and the volume of new research, in comparison to other systems approaches.} area of systems research; the barrier to entry from many disciplinary backgrounds is comparatively low, since discipline-specific jargon is minimised; and there is a largely positive perception of complex systems in the wider community, where there is awareness at all. The downside to this approach includes a proliferation of poorly defined buzzwords; a level of hype that is reminiscent of early AI; and widespread use of complex systems as a superficial metaphor. Of course this is not the only legacy of the SFI: it has established a dynamic network of talented systems researchers affiliated with many Universities, which interacts with a network of powerful corporations (both networks are almost entirely confined to the USA), within a unique and effective business model. It has also published a high quality series of proceedings from invite-only workshops on a diverse array of interdisciplinary topics that have added considerable substance to the complex systems approach.

The New England Complex Systems Institute (NECSI) has taken a much more systematic approach to complex systems. Bar-Yam's \cite{Bar97} complex systems textbook collated a large survey of the mathematical techniques in complex systems, including iterative maps, statistical mechanics, cellular automata, computer simulation, information theory, theory of computation, fractals, scaling and renormalisation. These techniques were then further developed and applied to the brain, protein folding, evolution, development, and human civilisation, to demonstrate the interdisciplinary breadth of complex systems applications. In addition to publishing the only comprehensive textbook on complex systems\footnote{Other more specialised books do now exist, such as Boccara's \cite{Boc04} textbook on the physics of complex systems.}, NECSI hosts the International Conference on Complex Systems, which is the premier complex systems conference.

Taking these two organisations as representative of the complex systems approach, complex systems is essentially a refinement of the GST/cybernetics research agenda. There has, in fact, been remarkably little interaction between complex systems and second order cybernetics, soft systems or critical systems: the insights from these alternative systems approaches are neither acknowledged nor addressed. (GST is usually just not acknowledged -- the term never appears in \cite{Bar97, Gel94, Kau93, Sha06, Wal92, Lew92, Lan03}.) Nor do the contemporary soft approaches draw on the techniques of complex systems. This could be explained by a general aversion to mathematics -- for example, Midgley's four volumes of systems thinking explicitly aimed to exclude ``papers that were heavily dependent on a high level of mathematical reasoning'' \cite[p. xxi]{Mid03}. Consequently, of 76 papers on systems thinking, only one high-level paper on complex systems was included. There has been greater interaction between complex systems and other `hard' systems approaches: OR and systems analysis share techniques such as genetic algorithms \cite{Gol89} and particle swarm optimisation \cite{KE95} with complex systems, while the new area of complex systems engineering \cite{BMB06} augments systems engineering problems with complex systems approaches. To show just how closely the aims of complex systems overlap with the GST/cybernetics research agenda, analogs of the ideas introduced in Sections \ref{secGST} and \ref{secCybernetics} are now identified within complex systems.

Whereas von Bertalanffy imagined unity of the sciences through isomorphisms, Gell-Mann \cite[p. 111]{Gel94} talks of building staircases (or bridges) between the levels of science, while Bar-Yam \cite[p. 2]{Bar97} cites universal properties of complex systems, independent of composition. Self-organisation, which has been labelled antichaos \cite{Kau91} due to its stabilising dynamics, is a substitute for equifinality. However, equal attention is now given to chaotic dynamics -- which exhibit sensitivity to initial conditions -- to demonstrate the limits of the classical state-determined system. Goal-directed behaviour is studied under the heading of autonomous agents \cite[p. 49]{Kau00}. Organisation is still of central importance \cite[p. 81]{Kau00} -- especially self-organisation \cite{Kau93, Sha01}, \cite[p. 691]{Bar97} -- which largely follows in the tradition of either Prigogine \cite{NP77} (viewed as an open system far from equilibrium) or Ashby \cite{Ash62} (viewed as an increase in organisation or an improvement in organisation).

The preoccupation in complex systems with quantifying the complexity of a system is a minor extension of the desire to understand organisation in general, and qualitatively expressed in Weaver's concept of organised complexity. The most common way to quantify complexity is by the amount of information it takes to describe a system, which follows from the use of Shannon information theory in cybernetics, but also borrows from theoretical computer science the notion of algorithmic complexity. Ashby's conceptualisation of adaptation and learning as an adaptive walk in the parameter space of a dynamical system is followed by Kauffman \cite[p. 209]{Kau93}, and generalised theories of adaptation and evolution have been developed by Holland \cite{Hol95} and Bar-Yam \cite[p. 531]{Bar97} respectively. Meanwhile, genetic algorithms \cite{Gol89}, artificial neural networks \cite{Hop82}, reinforcement learning \cite{SB98} and ant colony optimisation \cite{DMC96} are a selection of biologically inspired practical techniques used to generate adaptation in complex systems.

Von Neumann's \cite{von66} theory of natural and artificial automata are still studied in a mostly pure mathematical branch of complex systems -- cellular automata -- that can be expected to gain increasing attention from potential applications in areas such as Field Programmable Gate Array (FPGA) design and nanotechnology. Emergence continues to be a central concern in complex systems \cite{Hol98, Dar94, Bar04}. Boulding's hierarchy among the disciplines of science is maintained by Gell-Mann \cite[pp. 107-120]{Gel94}, although Simon's view of systems as nested hierarchies has been mostly supplanted by the consideration of systems as networks \cite{BA99, WS98}. An approximate translation between GST/cybernetics and complex systems is summarised in Table \ref{tabGSTCS}.

\setlength\extrarowheight{4pt} 
\begin{threeparttable}
\caption{Comparison of the themes of general systems theory and cybernetics with complex systems.} \label{tabGSTCS}
\begin{tabular}{|>{\raggedright}p{0.45 \textwidth}|>{\raggedright}p{0.45 \textwidth}|} \hline

\textbf{General Systems Theory} & \textbf{Complex Systems} \tabularnewline
\textbf{and Cybernetics} &  \tabularnewline \hline
Unity of science through isomorphisms & Coherence of science through bridges\tabularnewline \hline
Isomorphic mappings & Universality classes \tabularnewline \hline
Emergence & Emergence \tabularnewline \hline
Organisation & Self-organisation \tabularnewline \hline
Organised complexity & Complexity \tabularnewline \hline
Adaptation & Adaptation/evolution \tabularnewline \hline
Equifinality & Chaos and antichaos\tabularnewline \hline
Goal-directed behaviour & Autonomous agents \tabularnewline \hline
Automata & Cellular automata \tabularnewline \hline
Hierarchies & Networks \tabularnewline \hline
\end{tabular}
\end{threeparttable}

\vspace{\baselineskip}

With so much commonality, what is different about the field of complex systems? In terms of the research agenda, \emph{very little}.

Phelan \cite{Phe99} notes the broad similarities between complexity theory (referred to here as complex systems) and systems theory (by which Phelan means systems methodologies in operations research, engineering and management science, following from the pioneering systems work of von Bertalanffy, Ashby and Boulding), but attempts to also distinguish between the two. According to Phelan, systems theory is predominantly focused on intervention, whereas complex systems is more interested in exploration and explanation. Whereas the system dynamics models of systems theorists only capture the aggregate flow of quantitative parameters, complex systems models consist of unique individual agents capable of learning, planning and symbolic reasoning, capturing the micro-diversity that systems theory did not. Thirdly, Phelan claims that `complexity' has a different interpretation within the two fields. For systems theorists, ``complexity is a function of the number of system components and the amount of interaction between them''. In contrast, in complex systems complexity ``is something of an illusion---it is something that \emph{emerges} when several agents follow simple rules''.

Of these three distinctions, the second is the the most meaningful. The first distinction is a matter of emphasis: both systems theory and complex systems are in the business of both explanation and intervention (consider the interventions recommended by the companies that comprise the Santa Fe Info Mesa). The third distinction may be true for the work of some complex systems theorists -- such as Holland \cite{Hol95, Hol98} and Wolfram \cite{Wol02} -- but does not hold for others, such as Bar-Yam \cite{Bar97} and Kauffman \cite{Kau93}. It is more accurate to observe that there are many more ways to think of complexity (algorithmic complexity, behavioural complexity, effective complexity, multiscale complexity, physical complexity and structural complexity, to name a few) in complex systems than the approach initially taken by systems theorists.

The principle difference between complex systems and GST/cybernetics is not the questions that are being asked, but the way in which they are answered. The techniques of complex systems have made substantial ground on the questions that the early systems theorists asked, but did not have the methods to answer. Agent based models, as identified by Phelan, are one component of this expanded toolkit within complex systems. It is only when one delves beyond the level of metaphor that complex systems reveals its contributions to the systems approach.

The analytical techniques within complex systems to understand organisation can be separated by three broad aims: to analyse the patterns, the scales and the dynamics of a system. Patterns that are discovered\footnote{Whether the patterns are a property of the system, the observer, or the relationship between them is a non-trivial question that is intimately related to the theory-ladenness of observation. See Dennett \cite{Den91} for an insightful discussion on patterns.} in measurements or descriptions of a system indicate the formation of structure in either space or time. Kauffman's boolean network models of self-organisation \cite{Kau93}, network theory, \cite{BA99, WS98} and computational dynamics \cite{Cru94} are examples of the analysis of pattern formation, where couplings, interactions and correlations respectively between parts of a system give rise to measurable regularities.

The relationship between an observer and a system can be analysed in terms of the scale and scope over which the observer interacts with the system. Multiscale analysis techniques \cite[p. 258]{Bar97, Bar05} provide a method for relating the possible system configurations that are observed at different scales. The development of the renormalisation group\footnote{Note that there is a connection between the renormalisation group and Simon's nearly decomposable systems \cite{Sim00}.} \cite{Wil79}, and dynamic renormalisation theory to analyse critical phenomena in physics provides an important general method of identifying the important parameters when the equations describing the system are independent of scale. In such cases, the method of separation of scales does not apply. Renormalisation operates by abstracting interdependencies between fine scale variables into interdependencies between coarse scale variables, and can be applied recursively. Fractals provide another significant set of techniques that enable multiscale analysis in complex systems, which were not available to GST.

Dynamics relates the behaviour of the system to its behaviour in the past. Dynamics can place constraints on future configurations, identify attractors and other patterns that occur in time. The calculus of nonlinear dynamical systems was not invented until after the peak of GST activity, meaning the early systems theorists talked of steady states and equilibria, rather than attractors and bifurcations. The dynamics of evolution and more generally adaptation (which includes learning as well as evolution) are of particular interest, as the only processes that are known to generate multiscale complexity (ie. increase organisation over multiple scales). Advances in biology since GST include a much better understanding of group selection pressures in evolution; swarm intelligence in social insect populations; and the operation of the adaptive immune system; all of which complex systems has drawn on and contributed to.

As well as analytical advances, the impact of the Information Technology revolution on complex systems can hardly be understated. Von Neumann's original automata were computed with pencil and paper. In contrast, Barrett's  TRANSIM Agent Based Model (ABM) (see for example \cite{Cas91}), running on Los Alamos National Laboratory supercomputers, has simulated the individual decisions of an entire city's motorists. Currently, there are ABM platforms capable of simulating over a million non-trivial agents, such as MASON \cite{LCPS04}. At its best, agent based modelling can combine all three threads of the complex systems approach to understanding organisation, simulating \emph{adaptive pattern formation over multiple scales}. While the computational power and relative simplicity of ABMs can be abused, acting as a substitute for thought\footnote{The complex systems community but also the wider scientific community has been particularly vocal in its criticism of Wolfram's \cite{Wol02} book on this account -- see for example \cite{Sha05a, Kra02, Kur02, Kad02, Gil02}.}, it has provided complex systems with a unique approach to model-building that stands in contrast to aggregate models, such as those used in system dynamics.

An example of an aggregate model that has been widely used in defence is the Lanchester differential equation for combat attrition, which relates the mass and effective firing rate of one side to the rate of attrition of the opponent. Rather than directly guess the functional form of aggregate attrition rates, the agent based approach models the individual agents, and the aggregate behaviour is computed from the `bottom up'. If the individual behaviour is easier to model than the aggregate behaviour, ABMs can eliminate the need to make simplifying assumptions about the macroscopic dynamics, instead allowing macroscopic patterns to `emerge' from the microscopic rules and interactions \cite{Bon02}. ABMs are also useful when the spatial configuration is relevant\footnote{If the spatial dimension is abstracted away but the interactions are retained, an ABM essentially becomes a network model, which has a set of analytical techniques that are increasingly popular within complex systems.} (c.f. Angyal above); when autonomy is distributed across many agents; when thresholds, nonlinearities or logical conditions govern individual decision-making; and when the rules change or the agents learn.

Of course, complex systems is not the only field to simulate the behaviour of individual agents in order to predict aggregate behaviour. In contrast to physics-based constructive simulations, the agents in ABMs are relatively abstract.  However, the relationships are typically richer, which gives rise to collective properties in addition to the properties of the simple agents.



\section{Defining `System'} \label{secDefnSystem}

\begin{quote}
\emph{A System is a set of variables sufficiently isolated to stay discussable while we discuss it.}
\begin{flushright}
W. Ross Ashby
\end{flushright}
\end{quote}

The historical approach to understanding systems provided a rich context within which individual contributions to systems theory could be assessed. In this section, my aim is to distill this into a concise definition, which captures the essence of the systems approach. Every definition must choose a position along a tradeoff between generality and depth of insight: as Boulding \cite{Bou56} notes, ``we always pay for generality by sacrificing content, and all we can say about practically everything is almost nothing.'' In contrast to Definition \ref{defSys1}, my definition of `system' will not be a catch-all container. As a relatively strong definition of system, it will exclude some entities that are legitimately systems in the informal sense -- such as closed systems -- but which are not the focal subject of the systems approach. Recalling that the focal subject is ``organised complexity'', this excludes entities with trivial structure or trivial behaviour, such as a contained gas, or well mixed solution of chemicals at equilibrium. The important points to consider from the previous section are:
\begin{enumerate}
\item Systems are an idealisation;
\item Systems have multiple components;
\item The components are interdependent;
\item Systems are organised;
\item Systems have emergent properties;
\item Systems have a boundary;
\item Systems are enduring;
\item Systems effect and are affected by their environment;
\item Systems exhibit feedback; and
\item Systems have non-trivial behaviour.
\end{enumerate}
Burke \cite{Bur06} defines a system as ``an idealisation of an entity as a complex whole.'' This is an extremely compact definition that incorporates the first seven aspects of systems listed above. Importantly, it explicitly recognises that as an \emph{idealisation}, a system abstracts away many details of an entity (the map is not the territory). A system is not a physical object, it is a representation that stands in for an entity, and it is always constructed and used by an agent (or observer). The use of the word \emph{entity} indicates that the subject of the idealisation may equally be abstract or substantial. The word \emph{complex} implies \emph{multiple components}, \emph{organised} in a non-aggregative arrangement. Also, complexity implies \emph{interdependence} \cite[p. 12]{Bar97}. The system \emph{boundary}, its \emph{enduring} nature, and \emph{emergent properties} in addition to the aggregate properties of the parts are all implied by the term \emph{whole} in Burke's definition.

However, the final three aspects of systems are not directly implied by this definition. Eight requires a system to be open. There is a subtle point here that requires explication, due to the fact that the system boundaries are chosen by the observer. Any open system can always be reframed as closed by expanding the system boundaries to include its environment. When applied recursively this method achieves closure with certainty, since the Universe is closed to the exchange of energy, matter and information. However, if for every possible system/environment partitioning the system is closed, then its long term behaviour will be trivial. Consequently, a more accurate way to state this requirement is that there exists at least one system/environment partition such that the system is open.

Together, eight and nine imply ten, since an open system requires more information than the initial conditions of the system to predict its future behaviour, and combinations of positive and negative feedback between the system and its environment are the source of \emph{non-trivial behaviour}. This implication was made precise by von Foerster \cite{von72}, who distinguished between trivial and non-trivial machines. The former have invariant input-output mappings and determinate behaviour, whereas the latter have input-output relationships that are determined by the machine's previous output. Non-trivial machines may be deterministic, yet are still unpredictable. According to this distinction, \emph{feedback} is necessary and sufficient for non-trivial machines. I now present my definition of system for this thesis, which incorporates all ten aspects of a systems approach:
\begin{defn}[System (7)] \label{defSysFinal}
A system is a representation of an entity as a complex whole open to feedback from its environment.
\end{defn}

I define a system as a representation rather than an idealisation, because even though the meaning is similar, representation has a well defined theoretical basis. This definition is too restrictive as a commonsense conception of system, but it provides insight on both the nature of systemic problems and the value of the systems approach. Because a system is a representation, it can stand in for the entity it represents. Systems are useful idealisations that complement the traditional abstractions within scientific disciplines, such as the point masses, frictionless surfaces, and massless springs of physics. All representations are based on simplifying assumptions, and as a consequence there are limits to their application.

This connection can help to clarify when the systems approach should apply. By rephrasing the list of systems aspects above, one can derive a set of simplifying assumptions of non-systemic approaches which deserve careful examination. This line of reasoning follows closely from Churchman's question, and has been explored elsewhere. For example, Bar-Yam \cite[pp. 91-95]{Bar97} takes this approach to contrast complex systems with the assumptions of thermodynamics. Weinberg \cite{Wei01} also provides an extensive discussion of the simplifying assumptions of science, and the importance in systems science of understanding the limits of applicability of simple models. When one or more assumptions are violated, then the systems approach may provide useful insights and more appropriate models (but never complete understanding). The assumptions of non-systemic approaches include:
\begin{enumerate}
\item The system is closed;
\item Averaging over time and across individuals is valid;
\item Superpositionality holds;
\item Space can be ignored;
\item Local structure is smooth;
\item Different levels are independent;
\item Control is centralised; and
\item Causation is linear.
\end{enumerate}

The assumption of a closed system enables the design of reproducible experiments with predictable outcomes from the same initial conditions, and the powerful inevitability of increasing entropy obtains, meaning dynamics are irrelevant to long-term behaviour and equilibrium states dominate. However, under this assumption only a limited number of properties of open systems far from equilibrium can be explained. Classical statistics affords great simplification for modelling systems with minimally biased estimators that average over time series and across individual members of an ensemble. However, averaging over time destroys information about temporal structure, while averaging over individuals hides the unique properties of the individual.

Superpositionality means that the whole is the sum of the parts: interactions within the system can be evaluated by calculating the sum of the pairwise interactions between components. This greatly simplifies the modelling process, and is consistent with Occam's Razor, because it assumes the minimal additional information about the system of interest. However, it assumes commutativity and that only first-order dyadic interactions exist. Therefore, superpositionality only strictly applies to aggregates, not systems where the structure of interactions affects the system's properties, and where the forces on a component have indirect effects. Similarly, assuming that spatial arrangement can be ignored simplifies the analysis of well-mixed systems, but if there is significant spatial structure, the assumption of global mixing may give misleading results, as local interactions depart from the average behaviour given by uniformly mixed interactions.

When a system's behaviour is fairly trivial, then one can safely assume that at finer scales, the local structure becomes smooth \cite[p. 258]{Bar97}, and that different levels can be analysed in isolation. The former assumption allows the application of calculus, while the latter allows the technique of separation of scales, which averages over fast processes and fixes the slow processes to analyse the intermediate dynamics \cite[p. 94]{Bar97}. These assumptions do not hold universally. If infinitesimals diverge or correlations occur over all scales -- both of which occur at a second order phase transition -- new modelling techniques are required.

Assuming control is centralised improves modelling efficiency: most components can be ignored because they are inert in the control process, and control becomes sequential, synchronous and linear. However, if autonomy is distributed across many system components, its behaviour may be qualitatively different to the centralised model. Typically, a system with distributed control may be expected to be less efficient but more robust and adaptive. Finally, the assumption that causation is always best modelled by a linear chain between effects and prior causes, as viewed by Laplace's demon, may not always hold. An example von Foerster \cite{von92} gave in an interview with Yveline Rey is how the process of a person switching on a light can be much more efficiently described as how causes in the future, ``to have the room lit'', determine actions in the present, ``turn the switch now''. It also explains why if turning the switch on does not illuminate the room because the globe is blown, subsequent actions find an alternative path to the same end, such as ``draw the curtains''.

This list of assumptions is not complete, but a large proportion of systems research can be motivated in terms of developing alternative techniques for when these assumptions do not hold. Alternatively, systems theory may generalise alternative techniques when they arise within more specialised disciplines, seeking to organise available techniques not by discipline, but by their assumptions about the system's structure and dynamics. Thus, a good deal of attention is paid to surfacing modelling assumptions and discussing limits of applicability within any systems approach. However, where GST found it difficult to say much about systems in general, contemporary systems approaches have had more success in identifying significant features of classes of systems that share certain properties (e.g. chaotic systems), or specific mechanisms for their generation (e.g. genetic algorithms) \cite{Sim00}. The GST ambition to unify science has been supplanted by the narrower but more realistic task of forging interdisciplinary links between specific models within the disciplines.






\section{Conclusion}
The systems movement is an attempt to understand organisation, a concept that is trivialised in mechanism and remains un-analysed in holism. Organisation is a compound concept that incorporates both structure and dynamics. The result of over half a century of systems thinking is not a general theory of organisation, but a loosely connected set of techniques, where each technique contributes some insight on the temporal and spatial structures of organised complexity. Because no single technique provides a complete understanding of organisation, knowledge of the limits of applicability of individual techniques is central to any systems approach.

Complex systems is arguably the most active contemporary field of systems research. It follows in the spirit of general systems theory and cybernetics, although it is less concerned with rhetoric on the unity of science as the development of quantitative models that have interdisciplinary application. Systems analysis and systems engineering have longer and much richer experience with real world systems applications, but constitute a comparatively superficial commitment to systems theory. All of these approaches can be broadly characterised as hard approaches, concerned with the exact scientific application of mathematical models.

The deepest divide between contemporary systems approaches occurs between hard and soft methods. Soft systems methodologies take a pluralist stance, where a systems model is taken to say more about the modeller and their assumptions than the system of interest itself. Advocates of soft systems approaches claim that hard approaches to social systems can be dangerous, because they do not account for the special nature of self-conscious and free-willed humans. Due to the deep philosophical differences between hard and soft systems approaches, there exists little constructive dialog between these communities.

A system is a representation of an entity as a complex whole open to feedback from its environment. The utility of a systems approach derives from the critical examination of simplifying assumptions. This helps to make explicit the associated limits of applicability, such that revision of the appropriate assumptions can extend the application of scientific model-building. The revisions apply in general to reductionist assumptions that wholes do not have properties apart from the properties of their components, and in particular to linear thinking about causation, composition and control.

\bibliography{ryanThesis}

\end{document}